\documentclass[conference]{IEEEtran}
\IEEEoverridecommandlockouts
\usepackage{cite}
\usepackage{amsmath,amssymb,amsfonts}
\usepackage{algorithmic}
\usepackage{graphicx}
\usepackage{textcomp}
\usepackage{xcolor}

\usepackage{comment}

\def\BibTeX{{\rm B\kern-.05em{\sc i\kern-.025em b}\kern-.08em
    T\kern-.1667em\lower.7ex\hbox{E}\kern-.125emX}}
\begin{document}

% \title{Common Issues in Conversion Process of a Large-scale Distribution Network Data from OpenDSS into a Real-Time Phasor Model}
\title{A Novel Asymmetric Distributed Voltage Regulation Scheme to Enhance the Utilization of Distributed Energy Resources}

% \author{\IEEEauthorblockN{Madhura Sondharangalla, Nikhil Korada, and Raja Ayyanar}
% 	\IEEEauthorblockA{Arizona State University, Tempe, AZ, 85281, USA }}
\author{\IEEEauthorblockN{Madhura Sondharangalla}
\IEEEauthorblockA{\textit{School of Electrical, Computer} \\ \textit{and Energy Engineering} \\
\textit{Arizona State University}\\
Tempe, USA \\
mbhashit@asu.edu}
\and
\IEEEauthorblockN{Dan Moldovan}
\IEEEauthorblockA{\textit{School of Electrical, Computer} \\ \textit{and Energy Engineering} \\
\textit{Arizona State University}\\
Tempe, USA \\
damoldov@asu.edu }
\and
\IEEEauthorblockN{Raja Ayyanar}
\IEEEauthorblockA{\textit{School of Electrical, Computer} \\ \textit{and Energy Engineering} \\
\textit{Arizona State University}\\
Tempe, USA \\
rayyanar@asu.edu }
}

\maketitle

\begin{abstract}
Passive voltage regulation in distribution grids using distributed energy resources (DERs) is becoming the standard for feeder voltage management in both under-voltage and over-voltage scenarios. Their simple control, ease of implementation, and fast response make them ideal for distribution feeders with high-DER penetration. This paper proposes a novel distributed voltage control for DERs based on the voltage at the point of common coupling, inverter rating, and operating active power. The proposed control is designed to utilize feeder inverters efficiently to reduce the burden on the inverters in both under-voltage and over-voltage operations and increase the fairness in reactive power ($Q$) support in over-voltage operations. The proposed control is implemented in static and dynamic simulations of a real feeder to validate the performance. Results for the under-voltage scenario are also presented in the paper. 
\end{abstract}

\section{Introduction}

\par ``Net zero emission by 2050 scenario'' predicts 400\% increment in household solar units by 2030 \cite{IEA}. The expected increase is at least 190 GW of new solar power added to the grid each year following 2022. The widespread popularity of rooftop solar units not only helps to achieve a reduced carbon footprint and helps limit global warming \cite{gwarming}, but it also plays an essential role in supporting the increasing energy demand driven by the adoption of electric heating and electric vehicles \cite{ev_increase}. An increase in distributed energy resources (DER), such as rooftop solar, brings major challenges to the power distribution grid, which is designed for a radial power flow. Voltage rise is one such challenge that limits the rate of growth of distributed generation due to limitations of the grid \cite{V_rise}. 

\par Fast intermittent generation of solar power puts a strain on legacy voltage regulation devices such as on-load tap changers (OLTCs) and capacitor banks. Regulating the distribution feeder voltages through DERs is becoming widespread supported by standards such as IEEE 1547-2018  \cite{IEEE1547}, to perform grid support functions including reactive power (Q) support such as volt-VAr and active power curtailment when needed \cite{volt-var}. Passive distribution grid controls are popular due to their simple but effective control in regulating the voltage. They can operate upon the states within the DER, such as point of common coupling (PCC) voltage, in contrast to the total grid awareness required for active grid control. Even though the solution is not optimal with passive control, passive control thrives because of its fast response and economic viability in implementation. 

\begin{equation}
\label{eqnVV}
Q_{sup} =
    \begin{cases}
        0.44\;S_{rated} & V_{PCC} \leq V_1 \\
        (V_2- V_{PCC})\frac{0.44\;S_{rated}}{(V_2-V_1)}  &  V_1 < V_{PCC} < V_2\\
        0 &  V_2\leq V_{PCC} \leq V_4 \\
        -(V_{PCC}-V_4)\frac{0.44\;S_{rated}}{(V_5-V_4)}) &  V_4< V_{PCC} < V_5 \\
        -0.44\;S_{rated} &  V_{PCC} \geq V_5
    \end{cases}
\end{equation}

\par As mentioned, volt-VAr control shown in \eqref{eqnVV} is popular among both researchers and the utility due to its simple but effective voltage regulation, where $V_1$, $V_2$, $V_3$, and $V_4$ define the boundaries for the voltage intervals as shown in Fig.\ref{new_curves_fig}. $V_{PCC}$ is the point of common coupling voltage for DER and $S_{rated}$ is the DER rating. The volt-VAr control commands the inverter to support$Q$based on the DER's PCC voltage. The $Q$ value depends on the maximum and the minimum points set by the curve, the rating of the inverter, and the instantaneous PCC voltage.

\par Even though the volt-VAr control is effective in mitigating both overvoltage and undervoltage violations, its insensitivity to the operating active power of the inverter makes it unfair for inverters operating at low active power. This can be mitigated by modifying the curve and integrating active power into the curve. For example, controlling the DER power factor generates a $Q$ command based on DER's operating active power \cite{pf_control}-\cite{volt-pf}. This scheme focuses on fair $Q$ allocation in the overvoltage scenario where DERs with higher $P$ should support a higher share of system $Q$ requirements.  However, in the under-voltage scenario, the DERs with higher active power are already contributing to voltage support. Hence, fair $Q$ support allocation requires the DERs with lower $P$ to provide a higher share of $Q$ support.  The challenges arising from this requirement are addressed in this paper. The fairer $Q$ support from the DERs leads to enhancing the lifetime of the inverter \cite{lifetime}-\cite{reliability}.

\par In this paper, a novel $Q$ support curve is introduced to overcome the fairness limitation of the volt-VAr control and enhance the performance of the active power-based control through,
\begin{itemize}
    \item A linear relationship between the voltage, operating active power, and the $Q$ of the DER.
    \item Enhancing the operation in low voltage region ($V_{PCC}$ $<$ 1.00) through employing low-power-operating DERs. 
\end{itemize}

\section{Proposed Voltage - Reactive Power Control}

\par
\begin{figure}[t]
\centerline{\includegraphics[scale=0.32]{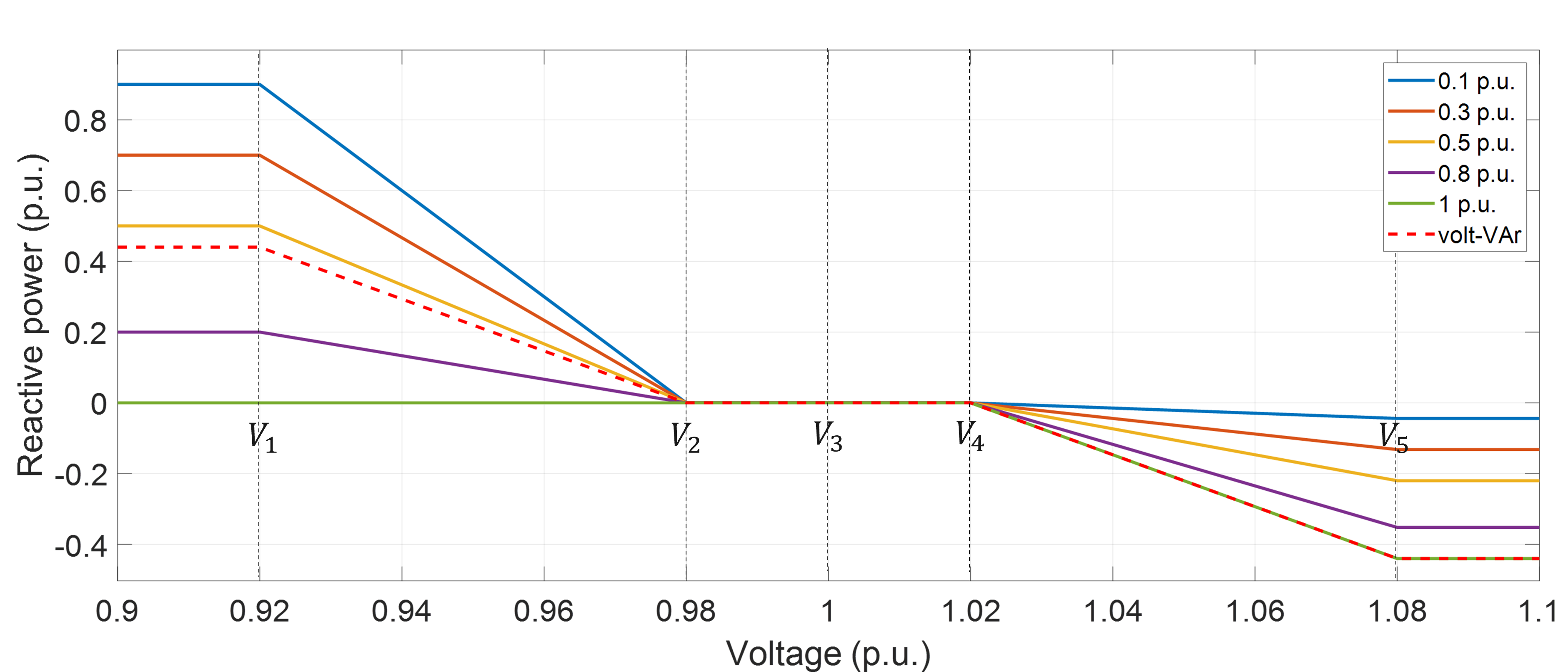}}
\caption{The relationship between voltage and $Q$ of an inverter under the proposed scheme compared with standard volt-VAr for a range of inverter active power.}
\label{new_curves_fig}
\end{figure}

 Characteristics of the proposed control follow a similar pattern to the volt-VAr control. As shown in \eqref{newcurve}, control is divided into five intervals based on the $V_{PCC}$. There are two saturation regions, one dead zone, and two droop regions. The two saturation regions are defined based on the operating active power, $P$, and the rated active power of the inverter ($P_{rated}$). The values are scaled based on the rating of the inverter ($S_{rated}$). The main difference between the proposed curve and the volt-VAr curve is that for a specific operating power, the proposed curve is not symmetric in under-voltage and over-voltage regions, as shown in Fig.\ref{new_curves_fig}.

{\small

\begin{equation}
\label{newcurve}
Q_{sup} = \\\frac{S_{rated}}{P_{rated}}
    \begin{cases}
       P_{rated} - P& V_{PCC} \leq V_1 \\
        (P_{rated} - P)\frac{(V_2- V_{PCC})}{(V_2-V_1)}  &  V_1 < V_{PCC} < V_2\\
        0 &  V_2\leq V_{PCC} \leq V_4 \\
        -\frac{0.44P(V_{PCC}-V_4)}{(V_5-V_4)} &  V_4< V_{PCC} < V_5 \\
        -0.44P&  V_{PCC} \geq V_5
    \end{cases}
\end{equation}
}

\par The droop curve on the over-voltage side ($V_{PCC}$ $>$ $V_4$) starts from zero at $V_4$ and reaches $-0.44S_{rated}\frac{P}{P_{rated}}$. As shown in Fig.\ref{new_curves_fig}, this results in a $Q$ support curve that scales with active power produced by the inverter. Between two inverters with similar $V_{PCC}$ and producing different levels of active power (partial shading or different orientation of PV panels), the inverter producing higher active power is commanded to support a higher $Q$ than the smaller one. The rated active power of the converter is defined as the maximum active power at which the inverter can operate while supporting the $0.44S_{rated}$ of Q.  

\par In the low voltage droop region ($V_{PCC}$ $<$ $V_2$), the injected $Q$ is based on the available active power of the inverter and defined by $\frac{S_{rated}}{P_{rated}}(P_{rated} - P)$. The saturation limit for this interval is different from that of the over-voltage region due to this difference. This modification creates an asymmetric reactive power control for the DER when injecting and absorbing Q.  The highlighted curve (Red) in Fig.\ref{new_curves_fig} emphasizes this difference when compared with an inverter operating at 30\% of the rated power. During the under-voltage interval, the peak injected $Q$ is 0.7$S_{rated}$; in the over-voltage region, the peak absorbing $Q$ is set to 0.132$S_{rated}$. 

\par Unlike in the over voltage scenario, in the under voltage case for fair $Q$ support, the inverters that operate at higher power levels support less reactive power compared to inverters operating at lower power levels. In a scenario where the voltage is low, inverters producing higher active power are already contributing to regulating the grid voltage by injecting active power. Therefore, to ensure fair regulation, inverters with lower power levels should support higher reactive power for regulation without overburdening the inverters with higher power levels. Since low-power inverters have a large reactive power buffer, this modification will not overburden the DER. Since injecting $Q$ is a function of $P$, it prioritizes the $P$ injection for the DERs. Allowing the DERs to inject reactive power beyond the limit of 0.44 Srated while still respecting the inverter VA rating reduces the overall reactive power needed from the distribution system, potentially eliminating the need for capacitor banks.

\begin{comment}

\par In the overvoltage region, the inverters that produce more active power are instructed to provide higher reactive power support than those producing less active power. This is done to counteract the behavior that worsens grid conditions by causing node voltages to increase. By giving priority to inverters with more active power, fairness is ensured in supporting voltage regulation for the feeder. This scenario is thoroughly explained in the research \cite{volt-pf} for high PV penetration conditions. Therefore, it is being neglected in this research due to repetitiveness. 
\end{comment}

\begin{figure}[h]
\centerline{\includegraphics[scale=0.43]{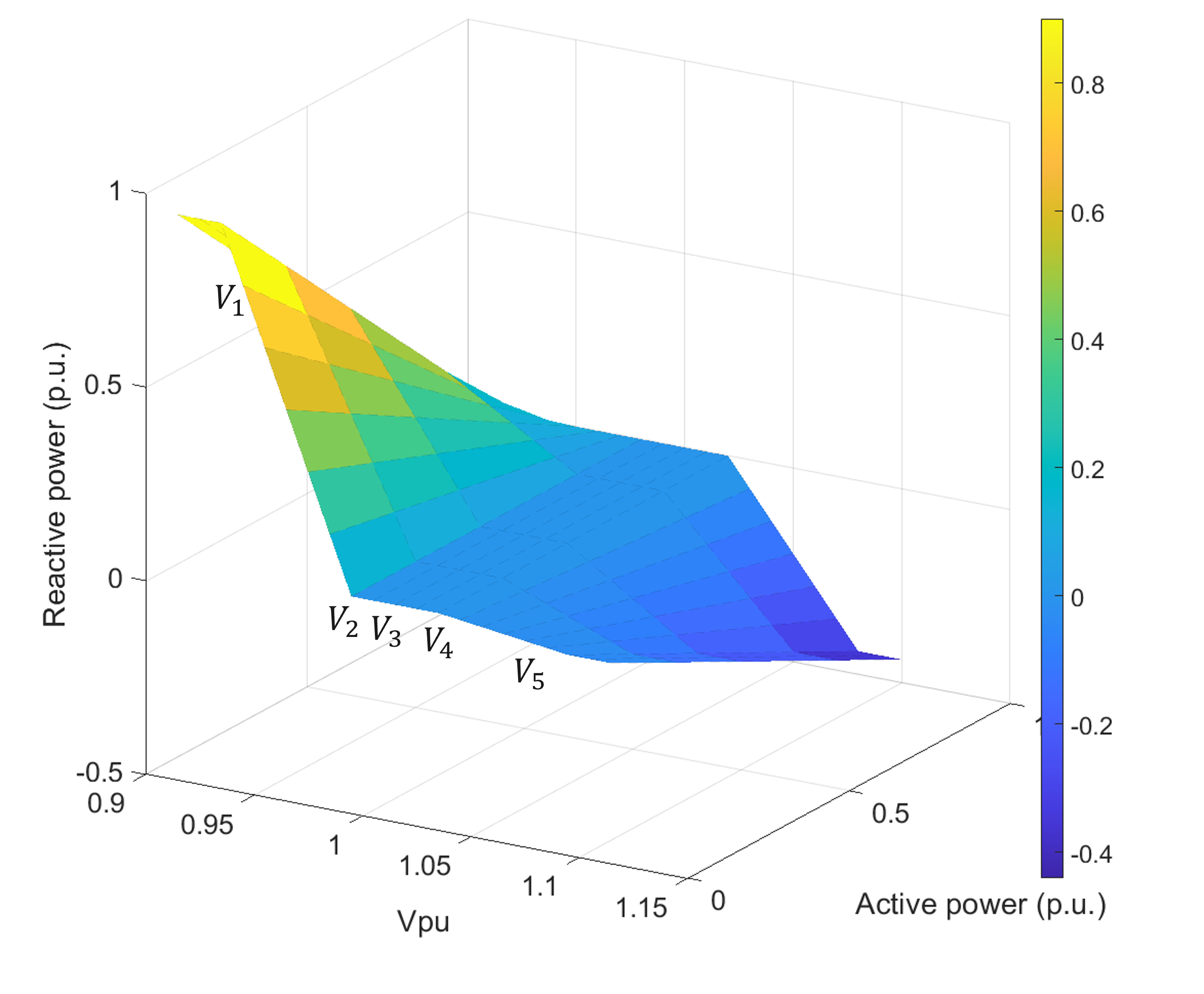}}
\caption{Three-dimensional view of the proposed reactive power support from DER as a function of DER active power and voltage.}
\label{new_curves_3d}
\end{figure}

\par Fig. \ref{new_curves_3d} shows the three-dimensional perspective of the relationship between DER's $Q$ support, $V_{PCC}$, and instantaneous DER active power in the proposed control mode. While $Q$ absorption at high voltages increases with higher active power, $Q$ injection at low voltages decreases as the active power increases.

\par Key benefits of the proposed control include,
\begin{itemize}
    \item Maintenance of the over-voltage regulation along the feeder by prioritizing DERs with higher active power to enhance $Q$ support.
    \item Linear relationship between voltage and $Q$ throughout the operating region. 
    \item Employing low-power DERs to share a higher burden in low-voltage scenarios to minimize potential solar power curtailment. 
\end{itemize}

\begin{figure}[h]
\centerline{\includegraphics[scale=0.4]{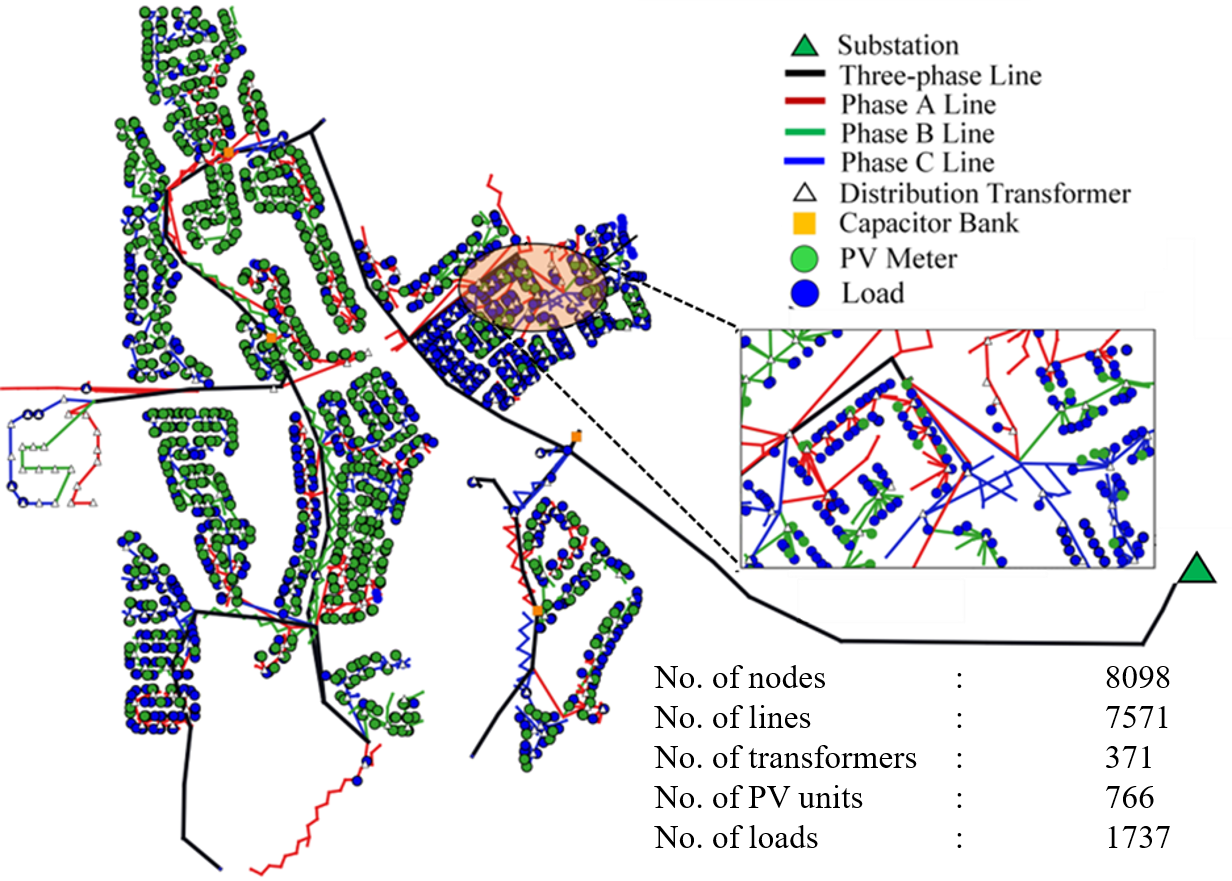}}
\caption{Distribution feeder selected to compare the effectiveness of the new DER-based voltage regulation mode}
\label{feeder}
\end{figure}

\section{Static Simulation Results}

\par The proposed distributed voltage regulation through DERs is implemented in a static simulation of a real distribution feeder to validate its performance in OpenDSS. The selected distribution feeder consists of 8098 nodes across 7571 lines, 371 transformers interconnecting 1737 loads, and 766 PV inverters, where the data is taken directly from a real distribution feeder located in Arizona.

\par  Since the feeder is located in Arizona, the July data set shows large load utilization (11.5085 MW), especially during the evening for air conditioners, which results in some node voltages dipping below 0.95 per. unit. In the March dataset, favorable weather conditions lead to low loads (no air conditioners), and high irradiance leads to high PV penetration and over voltage violations exceeding 1.05 per. unit. Generation of the model based on the AMI from the utility is explained in \cite{motaz}.

\par Since the static simulation consists of a 1-hour resolution data set, it results in discrete 24-hour static simulations. 1-hour resolution profiles include 1737 unique load profiles, which consist of active power and reactive power profiles, 766 PV profiles, and feederhead voltage profiles. Voltages of the feeder vary between 1.01 p.u. and 0.94  p.u. for the 24-hour period in July, where the minimum voltages occur at Hour 18 for the selected data set with several nodes reaching violation level ($\leq$ 0.95 p.u.).

\par The over-voltage condition and regulation follow the same concept of over-voltage regulations explained and demonstrated in \cite{volt-pf}. Therefore, the over-voltage condition and regulation are not included in this paper, and the focus will be on improving the under-voltage condition with a modified curve setting. The selected feeder features high load conditions in the July data set, specifically around Hour 18 and Hour 19. During these instances, the total PV generation is reduced due to the setting sun, but the high temperature in the area keeps the loads at near peak values.  

\begin{figure}[t]
\centerline{\includegraphics[scale=0.32]{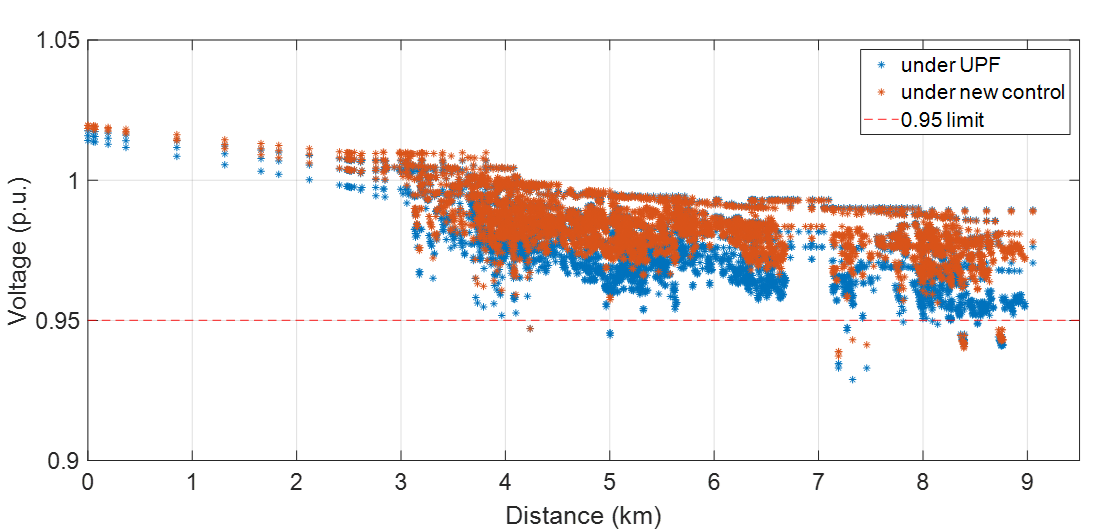}}
\caption{Voltage profile of the feeder at Hour 18 under UPF (blue) and proposed control (orange) obtained from OpenDSS static simulation}
\label{v_profile}
\end{figure}

\begin{figure}[t]
\centerline{\includegraphics[scale=0.35]{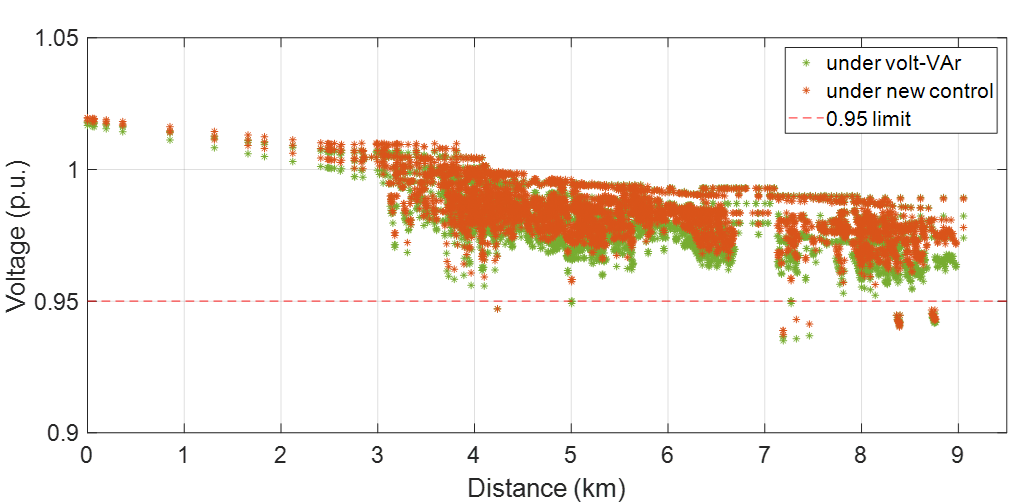}}
\caption{Voltage profile of the feeder at Hour 18 under volt-VAr IEEE 1547 default curve (green) and proposed control (orange) obtained from OpenDSS static simulation}
\label{v_profile_vv}
\end{figure}

\begin{figure}[t]
\centerline{\includegraphics[scale=0.31]{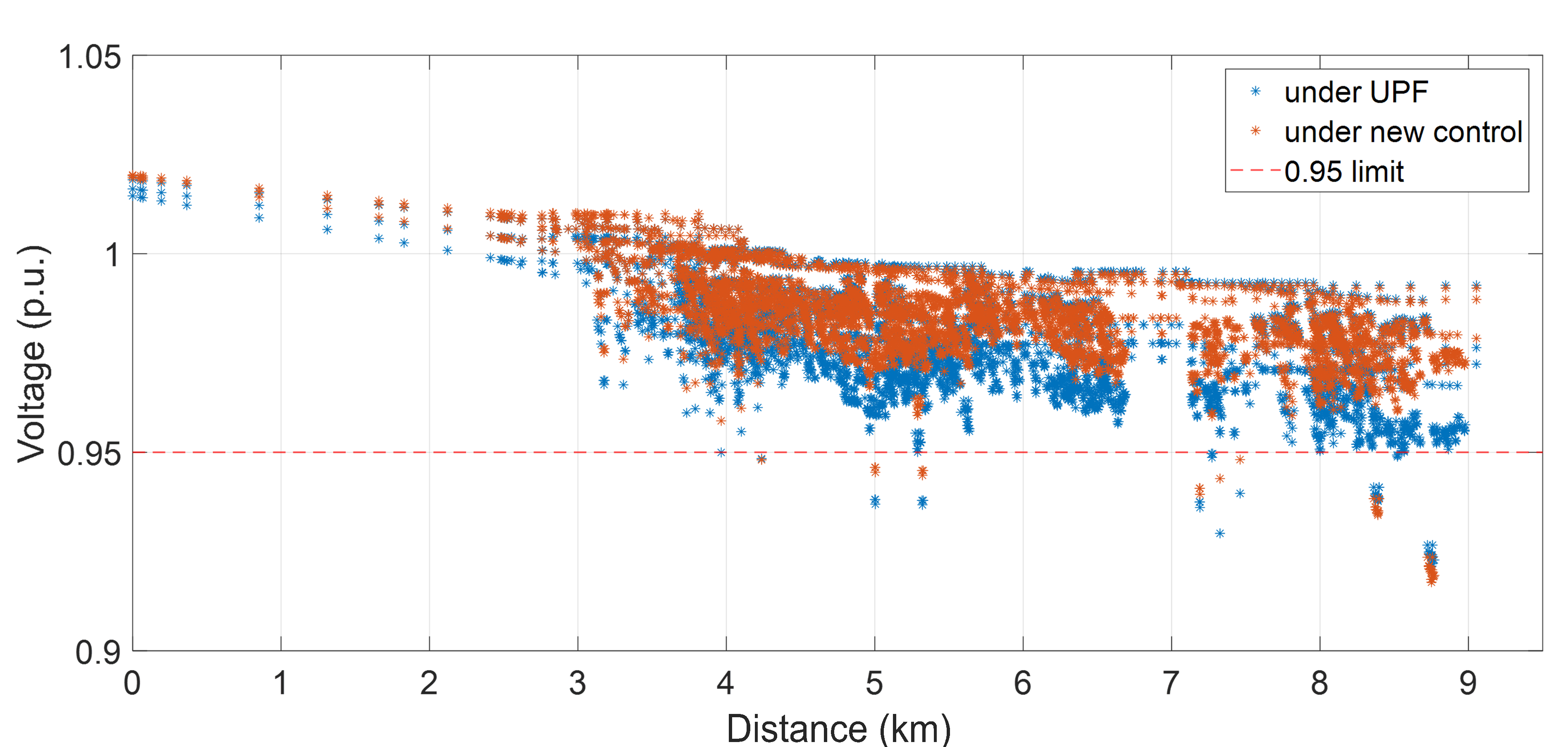}}
\caption{Voltage profile of the feeder at Hour 19 under UPF (blue) and proposed control (orange) obtained from OpenDSS static simulation}
\label{v_profile_19}
\end{figure}

\begin{figure}[t]
\centerline{\includegraphics[scale=0.32]{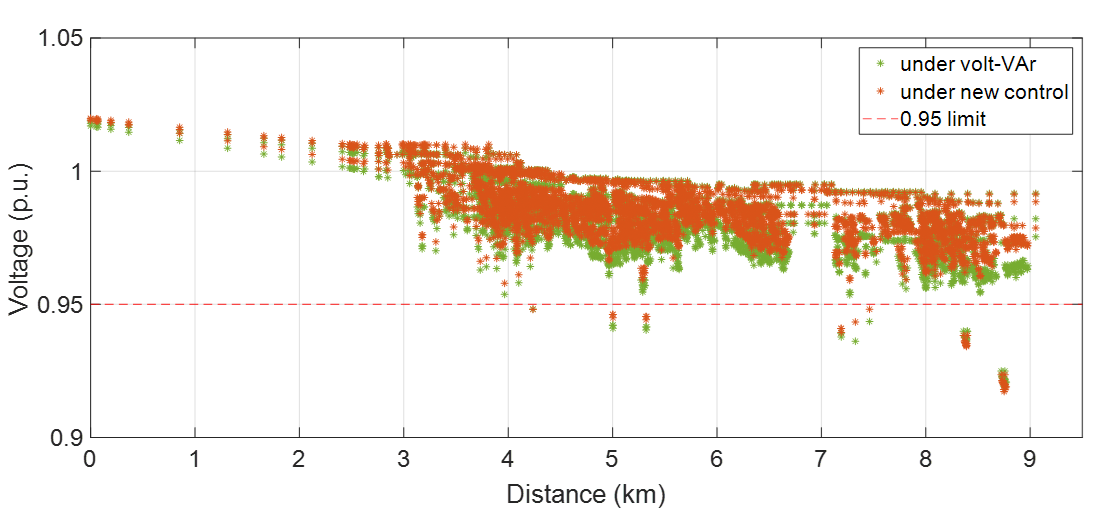}}
\caption{Voltage profile of the feeder at Hour 19 under volt-VAr IEEE 1547 default curve (green) and proposed control (orange) obtained from OpenDSS static simulation}
\label{v_profile_vv_19}
\end{figure}

{
\begin{table*}[t]
\centering
\renewcommand{\arraystretch}{1.5}

\caption{Node voltage statistics for Hour 18 snapshot simulation compared with no control, under volt-VAr and proposed control}

\begin{tabular}{|c|c|c|c|c|c|c|}
    
  \hline
  Control & Min. voltage& Max. voltage & Mean voltage & Total $Q$ from & Total $Q$ at the & Feeder\\
   type & (p.u.) & (p.u.) & (p.u.) &DERs (kVAr) & feederhead (kVAr) & losses (kW) \\
  \hline
  \textbf{UPF} & 0.9315 & 1.0179 & 0.9787 & 0 & 3207 & 423 \\
  & & & & & &\\
  \textbf{Volt-VAr} & 0.9375 & 1.0182 & 0.9813 & 370.38 & 2847 & 413.3 \\
  & & & & & &\\
  \textbf{Proposed control} & 0.9385 & 1.0192 & 0.9835 & 678.35 & 2528 & 408\\
  
  \hline

\end{tabular}
\label{node_voltage}
\end{table*}
}

{
\begin{table*}[t]
\centering
\renewcommand{\arraystretch}{1.5}

\caption{Node voltage statistics for Hour 19 snapshot simulation compared with no control, under volt-VAr and proposed control}

\begin{tabular}{|c|c|c|c|c|c|c|}
    
  \hline
  Control & Min. voltage& Max. voltage & Mean voltage & Total $Q$ from & Total $Q$ at the & Feeder\\
   type & (p.u.) & (p.u.) & (p.u.) &DERs (kVAr) & feederhead (kVAr) & losses (kW) \\
  \hline
  \textbf{UPF} & 0.92134 & 1.0186  & 0.9778 & 0 & 3105 & 446 \\
  & & & & & &\\
  \textbf{Volt-VAr} & 0.9192  & 1.0189 & 0.9807 & 450.74 & 2721 & 437 \\
  & & & & & &\\
  \textbf{Proposed control} & 0.9172 & 1.0208 & 0.9852 & 947.61 & 2154 & 426\\
  
  \hline

\end{tabular}
\label{node_voltage_19}
\end{table*}
}

\par Fig. \ref{v_profile} and Fig. \ref{v_profile_19} compare the voltage profiles of the feeder for the July data at Hour 18 and Hour 19 (blue) set that contains some under voltage violations. The proposed control is implemented for all the 766 inverters and commanded to regulate the feeder voltage. As OpenDSS is a static simulation, the proposed control is implemented as a Python script that calculates the required reactive power based on the node voltage, operating active power, and the rated power of the generator. With the proposed control, feeder voltage (orange) increases with the injection of $Q$ from the DERs. The new control mode performs better compared to the volt-VAr mode in voltage regulation in lower voltages, as shown in Fig. \ref{v_profile_vv} and Fig. \ref{v_profile_vv_19} for Hour 18 and Hour 19, respectively. As summarized in Table \ref{node_voltage} for Hour 18, the proposed control utilizes 1.8 times $Q$ to regulate the grid voltage due to the asymmetric curve. Therefore, the required $Q$ at the feederhead is also reduced by 26.8\% (3207 kVAr for UPF vs.2528 kVAr for proposed control) compared to UPF. A similar trend follows for the Hour 19 snapshot simulation, as summarized in Table \ref{node_voltage_19}, where the proposed asymmetric control injects 2.1 times higher $Q$ to regulate the voltage in low voltage conditions. There are 56 nodes (out of a total of 8098 nodes), as shown in Fig. \ref{v_profile_19} where the voltage with the new scheme is even lower than that of UPF and volt-VAr.  Potential cause for this could be the cross-phase effect where excessive Q injection in one phase can result in voltage drop in a different phase \cite{cross}.

\par The main difference between the two methods is in the utilization of the DER apparent power rating. The proposed method is designed to improve the utilization of low $P$ DERs for low voltage regulation depending on the voltage violation at PCC.

\section{Dynamic simulation results}

\begin{figure}[t]
\centerline{\includegraphics[scale=0.4]{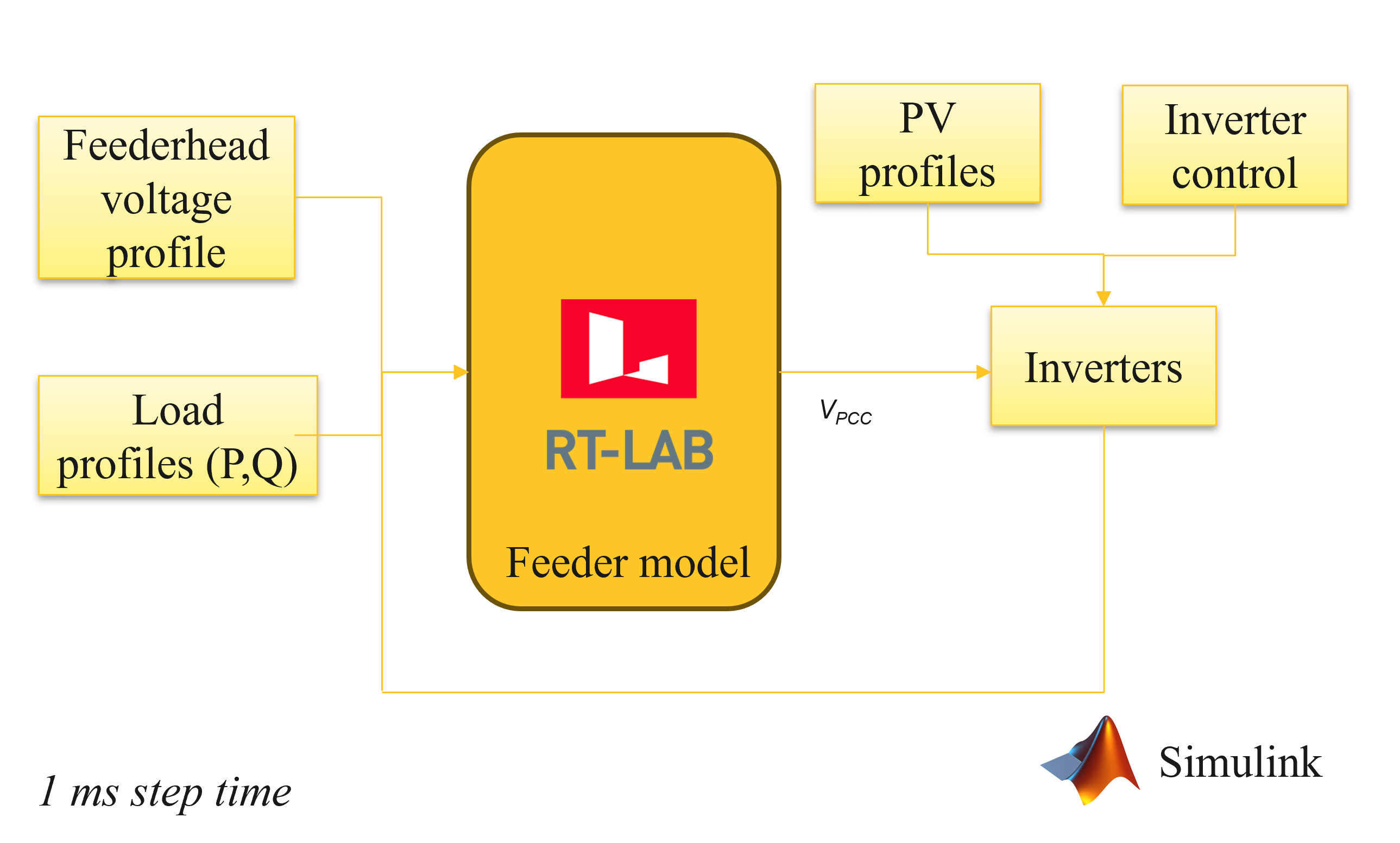}}
\caption{Simplified block diagram of the dynamic simulation model implemented in RT-Labs}
\label{sim_model}
\end{figure}

\par The same feeder model from the static simulation is used to verify the performance of the new control mode in the dynamic simulation. The feeder model is converted from the OpenDSS to ePHASORSIM model for the dynamic simulation. The static to dynamic simulation conversion process is explained in \cite{conversion}. The tested voltage magnitude error between the two models is less than 0.1 \% for 99\% of all the nodes, while the peak error is 0.44 \% between the two models. The typical error for the voltage angle is under 0.1 degrees for 99\%  of the nodes, with a maximum error of 0.25 degrees. 

\par Voltage, loads, and PV profiles contain time series data sets for feeder head voltage, active and reactive power for loads, and active power (maximum power point tracking) profile for the PV units as shown in Fig.\ref{sim_model}. Each time-series dataset has 1440 samples to create a 1-minute resolution profile for the elements. A high-resolution dataset was created using spline interpolation from the 24-hour profile simulation. The Opal-RT simulation is set to run at 1 ms step time with a linear interpolation to create the missing data. The dynamic simulation runs for 1440 seconds, emulating a 24-hour period in the feeder.  

\begin{figure}[t]
\centerline{\includegraphics[scale=0.32]{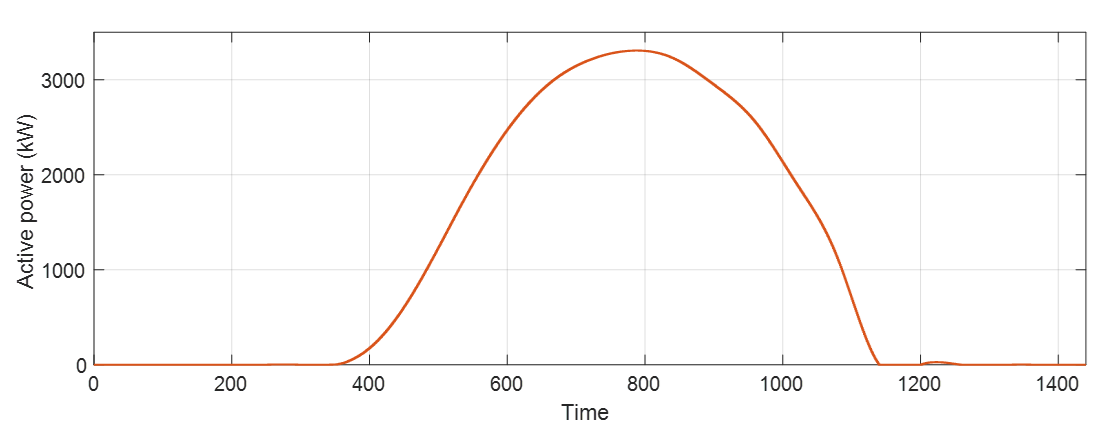}}
\caption{Total active power generated from 766 DERs in the selected feeder over a 24-hour period for July dataset}
\label{tot_P}
\end{figure}

\par In the simulation model, there are 766 Simulink models representing inverters, each connected to the feeder model at their respective nodes. Each inverter takes an active power command and the voltage of the node as inputs. At 798 seconds into the 1440-second simulation period, 766 DERs inject a peak total of 3.3 MW of active power into the grid. The DERs support the feeder from approximately 400 to 1322 seconds, as shown in Fig.\ref{tot_P}. The inverter control commands are also set as inputs to the inverter, with the inverter current feeding back to the node. The inverter control block commands the inverters to operate in their respective voltage regulation mode, which is predefined in the simulation. Inverters are set to operate in volt-VAr mode and the new control mode for 1440 s simulation time for comparison.

\begin{figure}[t]
\centerline{\includegraphics[scale=0.57]{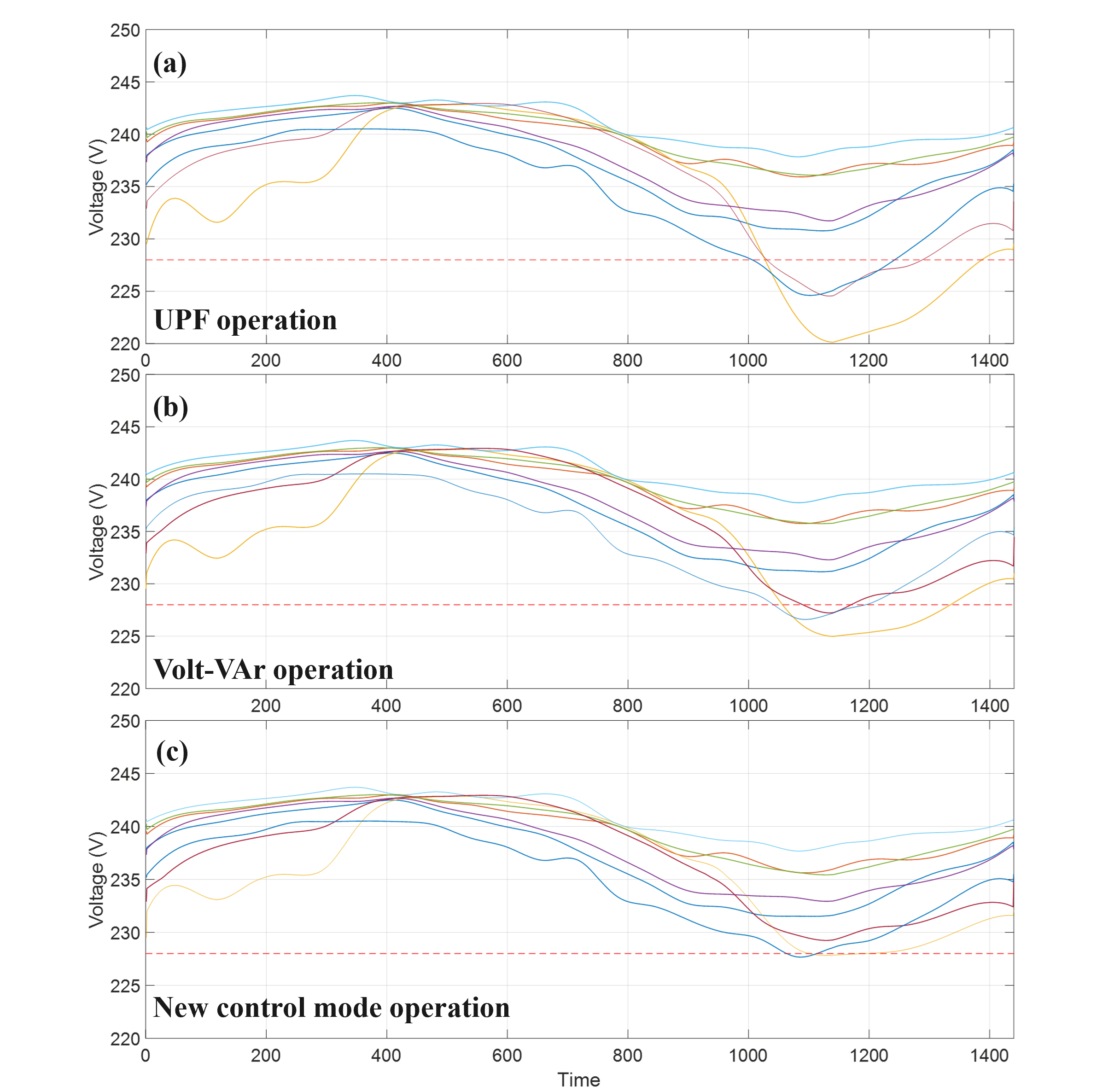}}
\caption{Voltages for selected feeder nodes under different DER operating modes. (a) unity power factor operation (b) Volt-VAr IEEE 1547-20218 operation (c) Proposed asymmetric control}
\label{min_v}
\end{figure}

\begin{figure}[t]
\centerline{\includegraphics[scale=0.44]{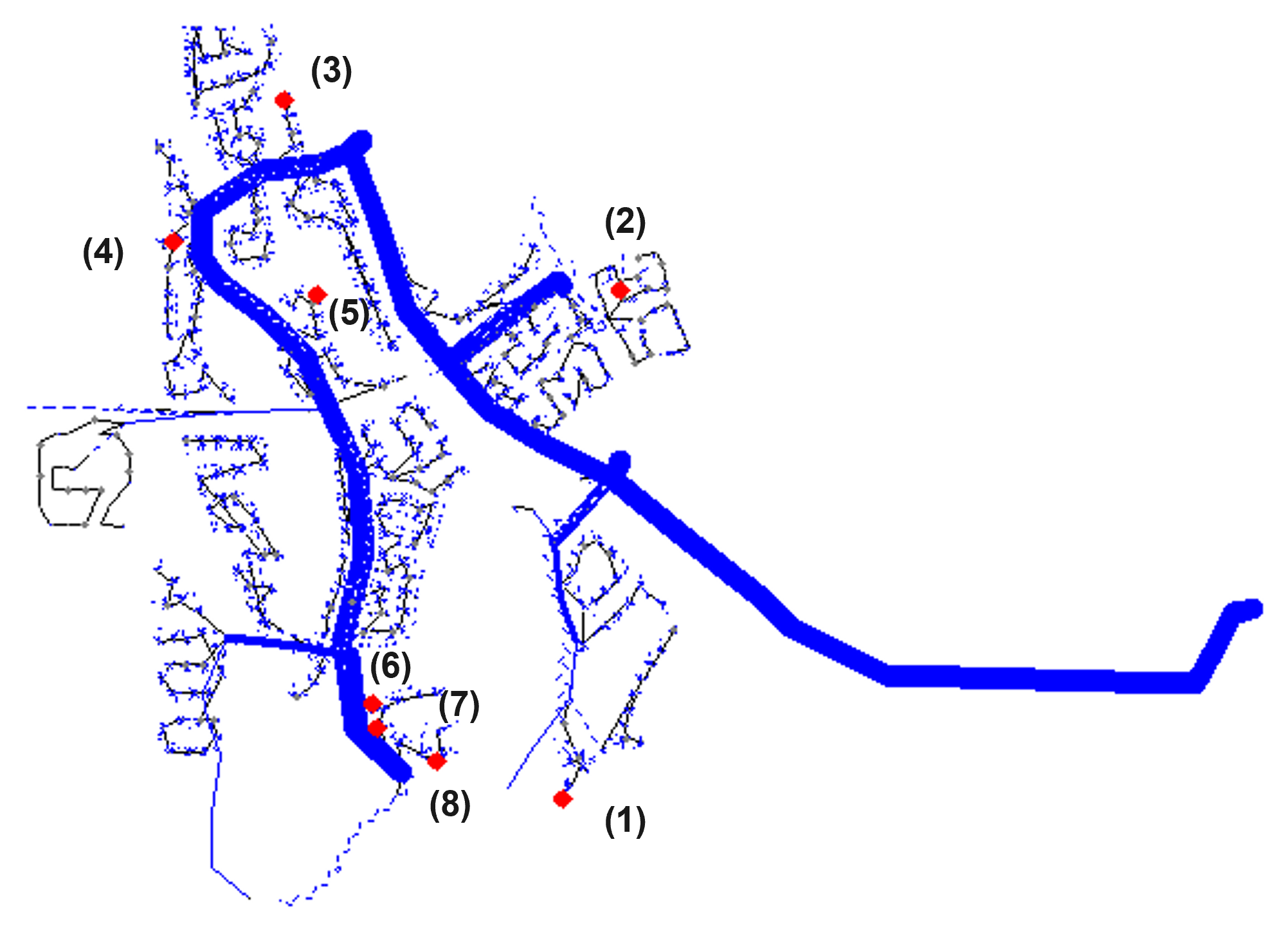}}
\caption{Locations of the selected nodes in the feeder showcased in the dynamic voltage results from the real-time simulation}
\label{node_place}
\end{figure}

\par Fig. \ref{min_v} illustrates the voltages of 8 selected feeder nodes under dynamic simulation. The selected nodes for the study are located throughout the feeder, as depicted in Fig. \ref{node_place}. The proposed new control mode performs better in regulating the minimum voltage than the volt-VAr during high load conditions due to the capability of using higher $Q$ from the DERs. As a result, the proposed asymmetric control eliminates the low voltage violations in the studied nodes. 

\par In over voltages, the new control mode regulates the feeder based on the active power generated from the DERs. The effectiveness of this operation is presented in the previous work, where it was shown to be able to regulate the grid voltage similar to volt-VAr with much less reactive power utilized. Note that since the $Q$ command is determined by the active power of the DERS, this control cannot provide $Q$ support during the night.  This issue has been addressed in the new work reported here by making $Q_{max}$ a function of ($P_{rated}$ - $P$) under low voltage conditions.

\par
\begin{figure}[t]
\centerline{\includegraphics[scale=0.31]{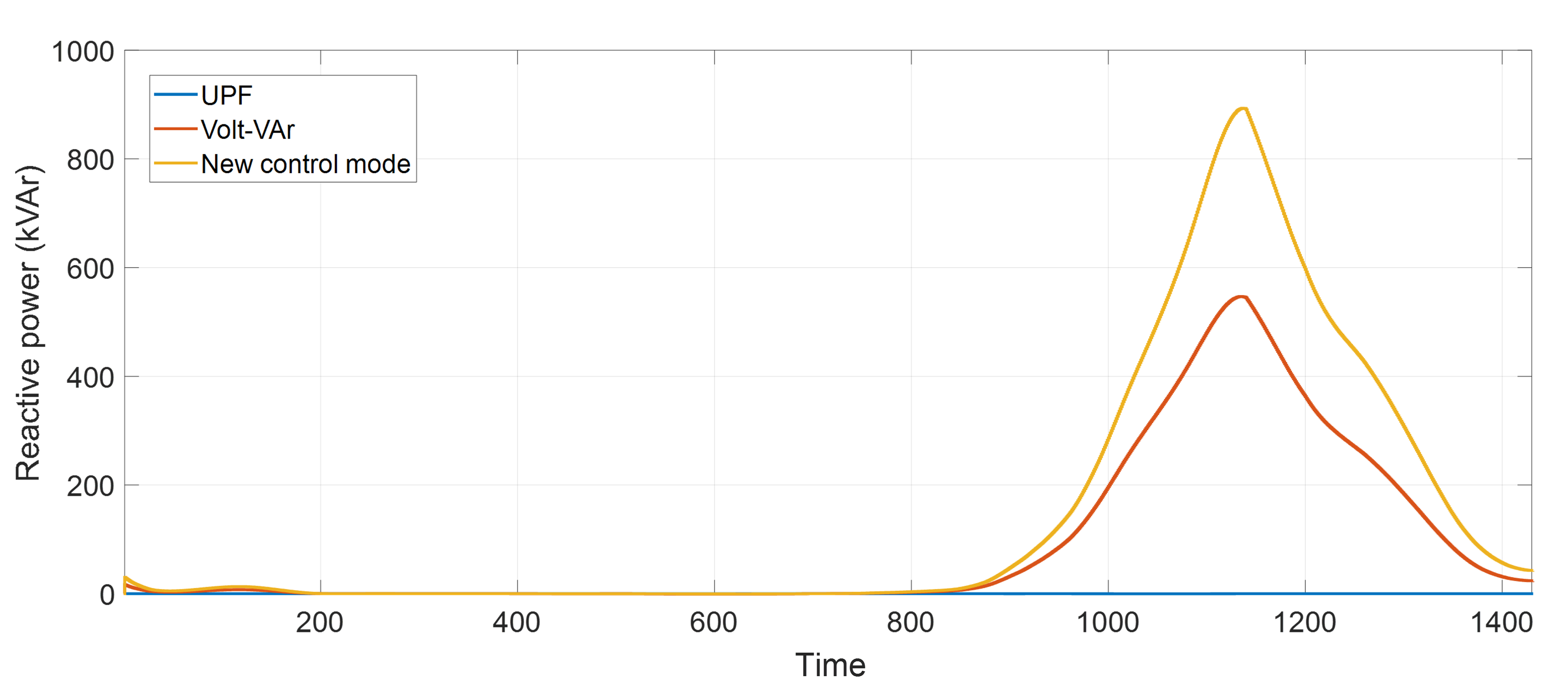}}
\caption{Total reactive power from the PV units over 24-hour cycle for different DER-based voltage regulation schemes}
\label{tot_react}
\end{figure}

 In under voltage cases, the new control mode utilized higher reactive power for regulating the grid voltage, as shown in Fig.\ref{tot_react}. During the same static simulation conditions, the new asymmetric control in dynamic simulation injects 69.4 \% more reactive power (672.9 kVAr for new control mode vs. 397.1 kVAr for volt-VAr at 1080 simulation seconds) for Hour 18 and 74.1\% higher $Q$ (897.7 kVAr for new control mode vs. 515.7 kVAr for volt-VAr at 1140 simulation seconds) for Hour 19. Higher portions of reactive power are supported by underutilized DERs due to their unique asymmetric curve, reducing the burden on high active power DERs. Small errors between the two models can be attributed to the modeling errors of the feeder and the profile interpolations. 

\par The proposed method not only ensures fairer $Q$ allocation but also reduces the total stress from the inverters for regulating the grid voltage in both over voltage and under voltage conditions. The proposed asymmetric control significantly reduces the feederhead $Q$ requirement by providing a much higher $Q$ for voltage regulation during the under voltage senarios. This further reduces the transformer $Q$ and the transformer loading for the feeder.

\section{Conclusion and Future Work}

The proposed asymmetric characteristics for DER reactive power as a function of the PCC voltage and the DER active power in both over-voltage and under-voltage regions ensure fair allocation of the burden of reactive power support from DERs operating at different active power levels. In the over voltage case, DERs with higher $P$ contribute more to the high voltage problem, and hence are required to contribute more $Q$ support (absorption) for voltage regulation.  This also results in lower combined $Q$ support from all the inverters compared to traditional volt-VAr control.  For the under voltage condition, DERs with higher $P$ are already contributing to raising the voltage through $P$ injection, hence, the burden for $Q$ support (injection) is allotted more to the DERs generating lower P.  Since DERs with lower $P$ have higher remaining capacity for $Q$ support within their VA rating, the proposed curve makes the injected Qmax to be also significantly larger for the low voltage case for effective feeder voltage management.  Static and dynamic simulation results verify the benefits of the proposed scheme.  In the under voltage condition, the new scheme also results in lower transformer loading and lower reactive power needed at the feeder head.

\end{document}